\documentstyle[12pt]{article}
\begin{document}
\parskip 10pt plus 1pt
\title{Duality in Equations of Motion from Spacetime Dependent Lagrangians}
\date{}
\maketitle
\centerline{\it Rajsekhar Bhattacharyya $^a$ and Debashis Gangopadhyay $^{b}$
\footnote{e-mail:debashis@boson.bose.res.in}}
\centerline {$^a$ Department of Physics, Jadavpur University, Calcutta-700032, INDIA}
\centerline{$^b$ S.N.Bose National Centre For Basic Sciences,}
\centerline {JD-Block,Sector-III,Salt Lake, Calcutta-700091, INDIA.}
\baselineskip=20pt

\begin{abstract}
Starting from lagrangian field theory and the variational principle, we show
that duality in equations of motion can also be obtained by introducing explicit
spacetime dependence of the lagrangian. Poincare invariance is achieved precisely
when the duality conditions are satisfied in a particular way. The same analysis
and criteria are valid for both abelian and nonabelian dualities. We illustrate
how (1) Dirac string solution (2) Dirac quantisation condition
(3) t'Hooft-Polyakov monopole solutions and (4) a procedure emerges for
obtaining {\it new} classical solutions of Yang-Mills (Y-M) theory. Moreover,
these results occur in a way that is strongly reminiscent of the {\it holographic
principle}.


\end{abstract}
\newpage
{\bf 1. Introduction}\\
Most theories with symmetries have their roots in lagrangian field theory as
derived from the variational principle. Recently, a new symmetry has attracted
attention,{\it viz.},duality. Considerable literature  exists where duality
has been studied in the above framework. For free Maxwell theory, this was
first done in a non-manifestly covariant approach by Zwanziger [1] and Deser
and Teitelboim [1]. Zwanziger's [1] lagrangian depended on a fixed four vector
and manifest isotropy was lost. This was regained when the electric and magnetic
charges fulfilled a quantisation condition. Deser and Teitelboim [1] constructed
duality transformations by a time-local generator and showed: free Maxwell action
and stress tensor components were duality invariant and the generator conserved;
in Y-M theory no transformation exists (at the level of fields $A_{\mu}$) that
gives the desired rotations and leaves the action invariant although stress tensor
was duality invariant. Deser [1] clarified that duality transformations are not
possible in nonabelian theories as these have minimal (self-) coupling.Action
description of the so-called self-dual boson fields in any even spacetime
dimension also uses this approach [2]. Manifestly Lorentz-covariant approaches
for bosonic fields using an infinite set of auxiliary fields exist in [3];
in  [4] actions with a finite set of auxiliary fields were used and the
covariant version of refs. [1b-2] constructed. In [5] the covariant version of
Zwanziger's action [1] was proposed and its connection to [4] shown in diverse
dimensions with and without sources. Coupling of the actions [1b-2] to arbitrary
external sources was done in [6]; in [7] this was solved for the covariant
approach of [4]. Based on [4], the covariant worldvolume action for the M-theory
super-5-brane coupled to the duality-symmetric D=11 supergravity was constructed
as well as the Lorentz-covariant action for Type IIB supergravity [8].
Models with duality-symmetric or self-dual fields are in [9] while other
approaches are in [10].

The motivation of the present work comes from the recent discoveries of the
behaviour of field theories at the boundaries of spacetimes [13].
Specifically, gauge theories have dual description in gravity theories in one
higher dimension. The theory in higher dimensions is encoded on the boundary
(which has a lower dimension) through a {\it different} (i.e. the particle spectrum
is different) field theory in a lower dimension. The operators in the lower
dimensional theory are now {\it composite}. This encoding is through
the phenomenon of {\it duality}, manifesting itself in relations between the
coupling constants of the two theories. This discovery of Maldacena and others
[13] is a concrete realisation of the {\it holographic principle} of t'Hooft [13]
according to which the combination of quantum mechanics and gravity requires the
three dimensional world to be an image of data that can be stored on a two
dimensional projection much like a holographic image.

In this context, we give another approach to electromagnetic duality and show
that some analogue of the holographic principle seems to exist even at length scales
far larger than that of quantum gravity.{\it This is the formalism of spacetime
dependent lagrangians coupled with Schwarz's view [11] that in situations with
fields not defined everywhere there exist exotic solutions like monopoles.}
(Note: $x_{\nu}$ dependence was already inherent in Zwanziger's work [1]).Such
solutions are related to duality. In this work we will be confined to
{\it classical solutions} of theories where {\it the fields do not couple to
gravity}. The $x_{\nu}$ dependence of the lagrangian will be embodied through
a function $\Lambda(x_{\nu})$, whose {\it finite} behaviour at {\it spatial
infinity} ${\bf x}=\pm\infty$ i.e. {\it boundary} (together with duality invariance
of equations of motion), gives exotic solutions ( Dirac string, Dirac monopole,
t'Hooft-Polyakov monopole etc.).

A field, by definition, is a quantity defined at all spacetime points. Fields
in a lagrangian must be defined everywhere. However, there are theories e.g.
the Dirac theory of monopoles [11] and	unified theories of strong,weak and
electromagnetic interactions where fields are defined only in a region. These
theories have monopole solutions and also duality invariance. We show that
these  solutions can also be understood by demanding the finite behaviour of
$\Lambda$ on the boundary at spatial infinity. Within the boundary, $\Lambda$
is like a {\it constant background external field and is non-dynamical}.
Hence it is ignorable. On the boundary, finiteness of $\Lambda$ encodes the
exotic solutions, restores Poincare invariance for the full theory and also
implies existence of {\it a new vector field} as a {\it classical} solution of
Yang-Mills theory. Both abelian and nonabelian cases are treated similarly.
{\it We stress that although our results occur at  length scales very much
larger than those of string theory (quantum gravity), some analogue of the
holographic principle still seems to exist.} The dynamics of $\Lambda$ on the
boundary and our results for the full quantum theories will be reported in
subsequent communications.

We first develop this formalism. Section 2 clarifies the role of $\Lambda$.
In Section 3 we show how  equations of motion themselves lead to Dirac-string like
configurations and how Dirac quantisation condition is obtained by
introducing a complex  interaction between  electric and magnetic charges.
In Section 4, t'Hooft-Polyakov monopole solutions [11] are obtained and a procedure
for obtaining {\it new} solutions to classical Y-M theory outlined. We also
show that the solutions can accommodate a new vector boson.
Section 5 lists the advantages of our method over those currently available.

Let the lagrangian $L'$	 be a function of fields $\eta_{\rho}$,
their derivatives $\eta_{\rho,\nu}$ {\it and the spacetime
coordinates $x_{\nu}$}, i.e. $L'= L'(\eta_{\rho},\eta_{\rho,\nu}, x_{\nu})$.
Variational principle [12] yields :
$$\int dV \left(\partial_{\eta}L'
- \partial_{\mu}\partial_{\partial_{\mu}\eta}L'\right) = 0\eqno(1)$$
Assuming a separation of variables :
$L'(\eta_{\sigma},\eta_{\sigma,\nu},.. x_{\nu})
=L(\eta_{\sigma},\eta_{\sigma,\nu})\Lambda(x_{\nu})$\\
($\Lambda(x_{\nu})$ is the $x_{\nu}$ dependent part and is a finite non-vanishing
function) gives
$$\int dV \left(\partial_{\eta}(L\Lambda)
- \partial_{\mu}\partial_{\partial_{\mu}\eta}(L\Lambda)\right) = 0\eqno(2)$$

{\bf 2. The Role of Lambda}\\
In this work we will be confined to {\it classical solutions} of theories
where {\it the fields do not couple to gravity}. Under these circumstances,

(1)$\Lambda$ is {\it not}  dynamical and is a finite, non-vanishing function
given once and for all at all $x_{\nu}$ multiplying the primitive
lagrangian $L$. {\it It is like an external field}, any allusion to the dilaton
is unfounded and equations of motion for $\Lambda$ meaningless.

(2)Duality invariance is related to finiteness of $\Lambda$. When fields are not
defined everywhere and equations of motion are duality invariant, finiteness
of $\Lambda$ on the spatial boundary at infinity leads to new solutions for
the fields. Requiring $\Lambda=1$ at $\bf x=\pm\infty$ gives back usual $L$
together with exotic solutions. Within the boundary $\Lambda$ is an ignorable
costant.

(3)Poincare invariance and duality invariance is achieved
through same behaviour of $\Lambda$. $\Lambda$ finite,but not a constant, gives
theories with duality invariance but not Poincare invariance.

(4)The finite behaviour of $\Lambda$ on the boundary {\it encodes the exotic
solutions of the theory within the boundary}. In this way we are reminded of
the holographic principle.

{\bf 3. The Dirac string-like field configuration and Dirac quantisation condition}\\
For a modified electrodynamics	$L'$ may be written as
$$L'= [-(1/4)F^{\mu\nu} F_{\mu\nu} + j^{\mu}A_{\mu}] \Lambda(x_{\nu})\eqno(3)$$
with $F^{\mu\nu} = \partial^{\mu}A^{\nu}-\partial^{\nu}A^{\mu}$
and $j^{\mu}$ a current. Dual of $F^{\mu\nu}$ is
$\tilde F^{\mu\nu} = (1/2)\epsilon^{\mu\nu\rho\sigma}F_{\rho\sigma}$. This
corresponds to $\bf E\Rightarrow\bf B$ and $\bf B\Rightarrow\bf - E$
in the field strength $F_{\mu\nu}$. Equations of motion obtained from $(2)$ are
$$\Lambda (\partial^{\mu}F_{\mu\nu})
+ (\partial^{\mu}\Lambda)F_{\mu\nu} - \Lambda j_{\nu} = 0\eqno(4a)$$
while the dual $\tilde F_{\mu\nu}$ satisfies
$$\partial^{\mu}\tilde F_{\mu\nu}=0\eqno(4b)$$
Duality invariance means identical equations of motion for $F$ and
$\tilde F$ ,i.e.
$\partial^{\mu} F_{\mu\nu}=0$. This implies
$$(\partial^{\mu}\Lambda) F_{\mu\nu} = \Lambda j_{\nu}\eqno(5)$$
There are two possibilities: (1)Finiteness of $\Lambda$ is assumed to be
independent of the behaviour of the fields and $\Lambda$ can be put equal to
the constant unity {\it a priori}. One then has {\it usual} electrodynamics and in
the absence of sources duality is present.(2)$\Lambda$ satisfies $(5)$ : $L'$
is a lagrangian whose equations of motion and Bianchi identity are invariant
under  duality rotations {\it even in presence of sources}. $\Lambda$ is
finite only if the fields behave in a certain way and this precisely
corresponds to the solutions mentioned above.

Now, at the theoretical level, neither of the equations
$\partial^{\mu}\tilde F_{\mu\nu}=0$ and $\partial^{\mu}F_{\mu\nu}=0$ are
more fundamental than the other.Remembering this let us discuss the second
possibility (i.e. $\Lambda$ satisfies $(5)$) by considering some specific
forms for $j_{\mu}$.\\
{\bf Case (a)}: Consider electrodynamics with only electric charge $e$.
Let
$j_{0}(x) = e \delta (x_{1}) \delta (x_{2}) \delta (x_{3})$;
$j_{i}(x) = 0$; \enskip $\Lambda(x_{\nu}) = \Lambda(x_{3})$.
$i$ runs over spatial coordinates.
Eq.$(5)$ then splits into two sets of equations : one for the
temporal index and another set of three	 for the spatial indices.
As $F_{3i}\not=0$, the second set gives a solution of $\Lambda$ as a constant
function of $x_{3}$. This solution when put in the first set gives
$F_{30}(0,0,0,x_{0}) = E_{3}(0,0,0,t) \Rightarrow \infty $.
$E_{3}$ is the electric field along the third direction. As this
is valid for all times, the solution is effectively time independent.\\
{\bf Case (b)}:Let
$j_{0}(x) = e \sum_{n=0}^\infty\delta (x_{1}) \delta (x_{2}) \delta (an+x_{3})$;
$j_{i}(x) = 0$; \enskip $\Lambda(x_{\nu}) = \Lambda(x_{3})$.
$i$ runs over spatial coordinates and $a$ is small and always positive.
Again, $\Lambda$ is constant and we get
$F_{30}(0,0,an,x_{0}) = E_{3}(0,0,an,t) \Rightarrow \infty$
as one possible configuration for the electric field. $E_{3}$ is again
time independent for reasons already mentioned.

As stated earlier, $\partial^{\mu} F_{\mu\nu}=0$ and $\partial^{\mu}\tilde F_{\mu\nu}=0$
must be placed on equal footing. So the just concluded analysis is also
valid for $\partial^{\mu}\tilde F_{\mu\nu}=0$. The only differences will be
(1) coupling $e$ (electric charge) replaced by	coupling $m$
(magnetic charge) (2) Maxwell's equations now have a corresponding magnetic
vector potential (3) Maxell's equations modified
with $div \bf B \not=0$ and $div \bf E=0$.The other Maxwell
equations will be accordingly modified. Consider case (a). Suppose we
remove the singular field right upto the origin.Then situation is similar
to Dirac construction of "infinite solenoid minus the string ".In the same
spirit case(b) reminds us of the  Dirac-string configuration. If now quantum
considerations of single-valuedness of magnetic wave function are imposed {\it
a la} Dirac, then Dirac quantisation conditions follow . This
is how the Dirac monopole solutions can be understood in our formalism.

Now consider a $U(1)\otimes U(1)$ gauge invariant theory. $A_{\mu}$,
$B_{\mu}$ are four-vector potentials corresponding to electric and magnetic
charges respectively; $F_{\mu \nu}$, $G_{\mu \nu}$ the respective field
strengths;  $j_{\mu}$, $k_{\mu}$  the electric and magnetic  (current)
sources with interactions between respective currents and potentials introduced
in usual way:

$$L_{1} = -(1/4) F^{\mu \nu}F_{\mu \nu} -(1/4) G^{\mu \nu}G_{\mu \nu}
+ j^{\mu}A_{\mu} + k^{\mu}B_{\mu}\eqno(6a)$$
\\$F^{\mu\nu}= \partial^{\mu}A^{\nu} - \partial^{\nu}A^{\mu}$ ;
\enskip	 $G^{\mu\nu}= \partial^{\mu}B^{\nu} - \partial^{\nu}B^{\mu}$ ;
\enskip	 $\tilde G^{\mu\nu}=(1/2)\epsilon^{\mu\nu\rho\sigma}G_{\rho\sigma}$.\\
$\partial^{\mu}j_{\mu}= \partial^{\mu} k_{\mu} = 0$ (current conservation);
$\partial^{\mu}A_{\mu}= \partial^{\mu} B_{\mu} = 0$ (transversality)\\
Therefore
$\partial^{\mu} F_{\mu\nu}= j_{\nu}\enskip ;
\enskip	 \partial^{\mu}\tilde F_{\mu\nu}=0 ;
\enskip \partial^{\mu} G_{\mu\nu}= k_{\nu}\enskip ;
\enskip	 \partial^{\mu}\tilde G_{\mu\nu}=0$.\\
Defining
$$\xi^{\mu\nu}=F^{\mu\nu}+\tilde G^{\mu\nu} ;
\enskip \beta^{\mu\nu}=\tilde F^{\mu\nu}-G^{\mu\nu}=\tilde \xi^{\mu\nu}\eqno(6b)$$
gives
$$\partial^{\mu} \xi_{\mu\nu}= j_{\nu}\enskip ;
\enskip	 \partial^{\mu}\tilde\xi_{\mu\nu}=-k_{\nu}\eqno(6c)$$
Note that for
$\xi_{\mu\nu}\rightarrow \tilde\xi_{\mu\nu}$ one has
$j_{\nu}\rightarrow -k_{\nu}$ ,
and for
$\tilde\xi_{\mu\nu}\rightarrow -\xi_{\mu\nu}$ one has
$k_{\nu}\rightarrow j_{\nu}$.
Contrast this to the usual case: for
$F^{\mu\nu}\rightarrow \tilde F^{\mu\nu}$, one has
$j_{\nu}\rightarrow k_{\nu}$
while for
$\tilde F^{\mu\nu}\rightarrow -F^{\mu\nu}$, one has
$k_{\nu}\rightarrow -j_{\nu}$;
with
$\partial^{\mu} F_{\mu\nu}= j_{\nu}\enskip ;
\enskip	 \partial^{\mu}\tilde F_{\mu\nu}=k_{\nu}$.
In the absence of $j_{\nu}$ and $k_{\nu}$ one gets back the usual case.

Here both the  $U(1)$ gauge fields ($A_{\mu}$, $B_{\mu}$) are independent. To
obtain a single independent field one can start with the fields as defined
in equation $(6b,c)$ and proceed like Zwanziger [1]. In fact, $6(c)$ is identical
to Zwanziger's starting equation ([1], eq.(2.1)) modulo a sign. The so-called
one independent field can be obtained as a {\it superposition} of two
independent fields--- each of which separately describes an
{\it electric-charge-only} world or a {\it magnetic-charge-only} world.
This is our view. From here we take a different route. We introduce a new
interaction between the electric and magnetic charges and invoke the
finiteness of $\Lambda$. The result is Dirac quantisation condition.

Consider a complex interaction $L_{2}'$ between the  electric
and magnetic charges via their respective four-vector potentials and (current)
sources:  $L_{2}'= i\alpha A^{\mu}B_{\mu} j^{\nu}k_{\nu}$.
$\alpha$ is a constant. In the classical theory the action has dimension
of angular momentum. It is then straightforward to verify that $\alpha$
has the dimension of {\it inverse} angular momentum i.e. $(\hbar)^{-1}$. Dirac in his
derivation of the charge quantisation using Maxwell's theory assumed that the
electron obeyed quantum mechanics. In our approach we are therefore motivated to
employ this semi-classical approach in the sense that we will be using a complex
interaction. If we take the lagrangian as
$$L'= L_{1} + L_{2}'\eqno(7)$$
one gets
$$\partial^{\mu} \xi_{\mu\nu}= (j_{\nu}+ic_{\nu}') \enskip ;
\enskip	 \partial^{\mu}\tilde\xi_{\mu\nu}=(-k_{\nu}-id_{\nu}')\eqno(8a)$$
with
$$c_{\nu}'= \alpha j^{\mu}k_{\mu}B_{\nu}\enskip;
\enskip d_{\nu}'= \alpha j^{\mu}k_{\mu}A_{\nu}\eqno(8b)$$
Here $j^{\mu}\rightarrow -k^{\mu}$ and $k^{\mu}\rightarrow j^{\mu}$ so that
$j^{\mu}k_{\mu}\rightarrow -j^{\mu}k_{\mu}$. Hence the duality transformation
gives
$$\partial^{\mu}\tilde \xi_{\mu\nu}=(-k_{\nu}-ic_{\nu}') \enskip ;
\partial^{\mu} \xi_{\mu\nu}= (j_{\nu}-id_{\nu}')\eqno(8c)$$
Comparing $(8c)$ with $(8a)$ we see that duality invariance cannot be obtained
starting from the lagrangian $(7)$.

Introducing $x_{\nu}$ dependence through $\Lambda(x)$,and putting
$L_{2}=f(\Lambda)L_{2}',\enskip c_{\nu}=f(\Lambda)c_{\nu}'$,
\enskip $d_{\nu}=f(\Lambda)d_{\nu}'$ the lagrangian is
$$L=L'\Lambda(x)= [L_{1} + L_{2}]\Lambda(x)$$
$$= [-(1/4) F^{\mu \nu}F_{\mu \nu} -(1/4) G^{\mu \nu}G_{\mu \nu}
+ j^{\mu}A_{\mu} + k^{\mu}B_{\mu}
+i f(\Lambda)\alpha A^{\mu}B_{\mu} j^{\nu}k_{\nu}] \Lambda(x)\eqno(9)$$
$f(\Lambda)$ a dimensionless function of $\Lambda$ such that
$$f(\Lambda)=0 \enskip or\enskip a\enskip finite\enskip constant$$ when $\Lambda=1$.

The conditions of duality invariance now become
$$\Lambda\partial^{\mu} \xi_{\mu\nu}
+[(\partial^{\mu}\Lambda  F_{\mu\nu}-\Lambda(j_{\nu}+ic_{\nu})]=0\eqno(10a)$$
$$\Lambda\partial^{\mu} \beta_{\mu\nu}
-[(\partial^{\mu}\Lambda  G_{\mu\nu}-\Lambda(k_{\nu}+id_{\nu})]=0\eqno(10b)$$
(Note that $\beta_{\mu\nu}=\tilde\xi_{\mu\nu}$). Duality invariance is obtained
if
$$[(\partial^{\mu}\Lambda  F_{\mu\nu}-\Lambda(j_{\nu}+ic_{\nu})]=0\eqno(11a)$$
and
$$[(\partial^{\mu}\Lambda  G_{\mu\nu}-\Lambda(k_{\nu}+id_{\nu})]=0\eqno(11b)$$
Let $\Lambda(x_{\nu})$ be a function of $x_{3}$ only and the sources
have only time components. So we have an electric charge at the origin and
a magnetic charge at $x_{3}=a$;
$\Lambda=\Lambda(x_{3})\enskip;\enskip
j^{i} = k^{i} = 0\enskip;\enskip
j^{0} = e \delta(x_{1}) \delta(x_{2}) \delta(x_{3})  ;
k^{0} = g \delta(x_{1}) \delta(x_{2}) \delta(x_{3}-a)$.
Then one has for $\nu=0,1,2$
$$(\partial^{3}\Lambda)	 F_{3\nu} = \Lambda(j_{\nu}+ic_{\nu})\eqno(12a)$$
$$(\partial^{3}\Lambda)	 G_{3\nu} = \Lambda(k_{\nu}+id_{\nu})\eqno(12b)$$
For $\nu=3$, $G_{33}=F_{33}=0$. So $c_{3}=d_{3}=0$. For $\nu=0$,solutions to $(12)$
are
$$\Lambda_{\infty} = \Lambda_{-\infty} exp[ e \delta(x_{1}) \delta(x_{2})/F_{30}(x_{1},x_{2},0,x_{0})]
exp[if(\Lambda)\alpha eg P_{0}(x_{1},x_{2},x_{3},x_{0})]\eqno(13a)$$
$$\Lambda_{\infty} = \Lambda_{-\infty} exp[ g \delta(x_{1}) \delta(x_{2})/G_{30}(x_{1},x_{2},0,x_{0})]
exp[if(\Lambda)\alpha eg Q_{0}(x_{1},x_{2},x_{3},x_{0})]\eqno(13b)$$
$$P_{0}(x_{1},x_{2},a,x_{0})
=(\delta(x_{1}))^{2}(\delta(x_{2}))^{2}\delta(a)B_{0}(x_{1},x_{2},a,x_{0}) / F_{30}(x_{1},x_{2},a,x_{0})\eqno(14a)$$
$$Q_{0}(x_{1},x_{2},a,x_{0})
=(\delta(x_{1}))^{2}(\delta(x_{2}))^{2}\delta(a)A_{0}(x_{1},x_{2},a,x_{0}) / G_{30}(x_{1},x_{2},a,x_{0})\eqno(14b)$$
$\Lambda$ must be finite. Let $\Lambda_{\infty} = \Lambda_{-\infty}=unity$.
Consider the set $(13a)$ and $(14a)$. Obviously the two
exponentials must reduce to unity. For the first exponential (as seen
in last section) {\it this corresponds to the Dirac string configuration
where $F_{30}\rightarrow\infty$ so that the exponential essentially becomes
unity}.For the second exponential, we see that in $(14a)$ the numerator has
singular $\delta-$functions and together with  $B_{0}\rightarrow\infty$ would
lead to a finite $P_{0}$ since $F_{30}\rightarrow\infty$. So second
exponential is $1$ if
$exp [if(\lambda)\alpha eg P_{0} ]=1$
i.e.$(exp[if(\Lambda)\alpha eg])^{P_{0}}=1$ (as $P_{0}$ is finite). Therefore
$$f(\Lambda)\alpha eg = 2 \pi n\eqno(15)$$

There are two possibilities:

(a)$f(\Lambda)=0$. Then the $U(1) \otimes U(1)$ invariance of $L_{1}$ is
not broken (from which Zwanziger's lagrangian can be retrieved via eq.$(6c)$)
and we just have  the Dirac string configuration (from the first exponential,
$F_{30}\rightarrow\infty$).

(b)$f(\Lambda)=\enskip a \enskip finite \enskip constant$. {\it Then the
$U(1) \otimes U(1)$ invariance of $L_{1}$ is broken; and putting
$\alpha=(\hbar)^{-1}$ we see that $(15)$ is just the Dirac quantisation
rule}.

So the Dirac quantisation rule is obtained by breaking the $U(1) \otimes U(1)$
invariance of the {\it unphysical} (i.e. two independent fields) lagrangian $L_{1}$.
For $\nu=1,2$ a similar analysis will again lead to $(15)$ and similarly
for $(13b)$ and $(14b)$.

{\bf 4. The t'Hooft-Polyakov Monopole Solutions}\\
Consider now a simple nonabelian generalisation of $(3)$:
$$L'= [-(1/4)G^{\mu\nu}_{a} G_{a\enskip\mu\nu}
+ j^{\mu}_{a}W_{a\enskip\mu} ] \Lambda(x_{\nu})\eqno(16)$$
$a,b,c$ are $SO(3)$ indices and $G^{\mu\nu}_{a}= \partial^{\mu}W^{\nu}_{a}
- \partial^{\nu}W^{\mu}_{a} - e\epsilon_{abc}W^{\mu}_{b}W^{\nu}_{c}$.
$j^{\mu}_{a}$ is an external current and $\tilde G^{\mu\nu}_{a}
=(1/2)\epsilon^{\mu\nu\rho\sigma}G_{a\enskip\rho\sigma}$.
Analogues of $(4a)$, $(4b)$ are:
$$\Lambda (D^{\mu}G_{a\enskip\mu\nu}) + (\partial^{\mu}\Lambda)
G_{a\enskip\mu\nu} - \Lambda j_{a\enskip\nu}=0\eqno(17a)$$
$$D^{\mu}\tilde G_{a\enskip\mu\nu}=0\eqno(17b)$$
Again duality invariance of the equations $(17)$ imply
$$(\partial^{\mu}\Lambda) G_{a\enskip\mu\nu} = \Lambda j_{a\enskip\nu}\eqno(18)$$
If we take $\Lambda=\Lambda(r)$, $j_{a \enskip i} = 0$ for all $i$ and $a$ and
$j_{a \enskip 0} = 0$ then one solution is a radial magnetic field
$\bf B_{a}= \bf H_{a}(r)$ with $\bf H_{a}(r)$ satisfying $D^{\mu}\tilde G_{a\enskip\mu\nu}=0$
and $D^{\mu} G_{a\enskip\mu\nu}=0$.

Now consider the Georgi-Glashow model with $L'$ defined as
$$L'= [-(1/4)G^{\mu\nu}_{a} G_{a\enskip\mu\nu}
+ (1/2)(D^{\mu}\phi)_{a}(D_{\mu}\phi)_{a}
- V(\phi)]\Lambda(x_{\nu})\eqno(19)$$
The gauge group is $SO(3)$, $G^{\mu\nu}_{a}$ is as defined before, and the
matter fields $\phi$ are in the adjoint representation of $SO(3)$. Equations of motion
are :
$$\Lambda (D^{\mu}G_{a\enskip\mu\nu}) + (\partial^{\mu}\Lambda)G_{a\enskip\mu\nu}
+ \Lambda e \epsilon_{abc} (\partial_{\nu}\phi)_{b}(\phi)_{c}
- \Lambda e^{2} \epsilon_{abc}\epsilon_{bc'd'}W_{\nu\enskip c'}
\phi_{c} \phi_{d'} = 0\eqno(20a)$$
$$(D^{\mu}D_{\mu}\phi)_{a}\Lambda + (D_{\mu}\phi)_{a}\partial_{\mu}\Lambda
= - (\partial_{{\phi}^{a}}V)\Lambda\eqno(20b)$$
and the Bianchi identities are:
$$D^{\mu}\tilde G_{a\enskip\mu\nu}=0\eqno(20c)$$
Duality invariance then leads to
$$(\partial^{\mu}\Lambda)G_{a\enskip\mu\nu}
= - \Lambda e \epsilon_{abc} (D_{\nu}\phi)_{b}(\phi)_{c}\eqno(21)$$
For $\Lambda=\Lambda(r)$ we have
$$\Lambda_{\infty}
= \Lambda_{0} exp[-e \int_{0}^{\infty} dr \left( (\epsilon_{abc}
(D_{\nu}\phi)_{b} \phi_{c})
(\partial^{i} r G_{a \enskip i\nu})^{-1}\right)]\eqno(22)$$
where $\Lambda_{p}$ is the value of $\Lambda$ at $r=p$;
$a,\nu$ are fixed; and there is a sum over indices $i,b$ and $c$.
$\Lambda_{\infty}$ must be finite. Choose this
to be the constant unity.This may be realised in the following ways :

{\bf(I)} $(D_{\nu} (\phi)_{b} \Rightarrow 0$, \enskip
$(\phi)_{c} \Rightarrow finite$, \enskip
and the product
$(D_{\nu}\phi)_{b} (\phi)_{c}$ falls off faster than $G_{a \enskip i\nu}$ for
large $r$. Then a constant value for $\Lambda$ is perfectly consistent with $(20b)$ and
the conditions become analogous to the Higgs' vacuum condition for
the t'Hooft-Polyakov monopole solutions where the duality invariance of the
equations of motion and Bianchi identities are attained at large $r$ by
demanding $(D_{\mu} \phi)_{a} \Rightarrow 0$ and $\phi_{a}\Rightarrow a\delta_{a3}$
at large $r$. {\it Note that our results are perfectly consistent with the
usual choice for the Higgs' potential $V(\phi)$ even though nothing has been
assumed regarding this}. Thus, the t'Hooft-Polyakov monopole solutions
follow naturally in our formalism.

{\bf (II)} $$\left(\epsilon_{abc} (D_{\nu}\phi)_{b}\phi_{c}\right) \Rightarrow 0\eqno(23)$$
and falls off faster than $G_{a \enskip i\nu}$ for large $r$ ($a$ and $\nu$ are fixed).
A solution is when
$$D_{\nu}  \phi = \alpha_{\nu} \phi$$
where $\alpha_{\nu}$ can be
(a) any Lorentz four vector field that is consistent with all the relvant equations
of motion and the finiteness of energy constraint.
(b) any Lorentz four vector field as in (a) but which may also carry
$SO(3)$ indices,i.e. $\alpha_{\nu}= \alpha^{a}_{\nu} \tau^{a}$,
$\tau^{a}$ being the generators of $SO(3)$, $a,b,c = 1,2,3$.
Putting this in $(20b)$ gives
($\Lambda$ now is unity)
$$[D^{\mu}(\alpha_{\mu}\phi)]_{a}= - (\partial_{{\phi}^{a}}V)\Lambda\eqno(24)$$
Problem therefore reduces to finding solutions of $(23)$ and $(24)$. These
will be discussed elsewhere. Eq.$(23)$ is like a {\it master equation} for
obtaining solutions of classical Y-M theory incorporating duality.

{\bf 5. Conclusion}\\
We have given an alternative way to understand duality using the approach of
spacetime dependent lagrangians coupled to the Schwarz postulate [11] that whenever
fields in a lagrangian are not defined everywhere one has monopole
solutions. Our results indicate that some analogue of the holographic principle
may be operative even at length scales far larger than the Planck scale in theories
which incorporate duality invariance in the equations of motion. The advantages
of our method are

(1)It gives a  procedure for obtaining {\it new} solutions to classical
Yang-Mills theory incorporating duality. In particular, the solutions
can accommodate a new vector field. It can be shown that none of the
known aspects of the Georgi-Glashow lagrangian are violated by the presence
of this solution. Details of such solutions and applications
to a full quantum theory will be discussed elsewhere.

(2)There is a possibility of a generalisation of the Bogomolny
equations (refer to equation $(23)$). Usual Bogomolny equations
seem to be the simplest first steps towards realisation of duality. This is
partly evident in that the Higgs' vacuum condition has been obtained
without specifying $V(\phi)$. Spontaneous symmetry breaking has neither been
invoked nor contradicted. $\Lambda$ is just required to be finite.

(3)The same analysis and criteria are valid for both abelian and
nonabelian dualities. {\it The Dirac-string solutions, the Dirac quantisation
condition and the t'Hooft-Polyakov monopole solutions all follow from the
same underlying principle}. This was not possible before [1b,1c].

(4)When $\Lambda$ is a constant, all known lagrangians exhibiting duality
become accessible. When $\Lambda$ is not a constant but some finite well
defined function everywhere, then a plethora of new lagrangians whose
equations of motion exhibit duality can be constructed (at the expense of
Poincare invariance).

(5)Finally, our method indicates that some flavour of the holographic principle
can be obtained in certain gauge theories {\it even at length scales very much larger
than those of quantum gravity}. This aspect has never been revealed before
in any study using standard methods.

We thank S.Deser for motivating clarification of many aspects
as well as providing references. We also thank A.Maznytsia for references
and the referee for important suggestions to improve the manuscript.
RB is supported by a CSIR-UGC SRF fellowship (no:2-21/95(I)/E.U.II).


\newpage
\begin{thebibliography}{15}
\bibitem [1] {kn:xx}D.Zwanziger, {\it Phys.Rev.} {\bf D3} (1971) 880;
S.Deser and C.Teitelboim, {\it Phys.Rev} {\bf D13} (1976) 1592;
S.Deser, {\it J.Phys.Math.Gen} {\bf A15} (1982) 1053.
\bibitem [2] {kn:xx}N.Marcus and J.H.Schwarz, {\it Phys.Lett.} {\bf 115B} (1982) 111;
R.Floreanini and R.Jackiw, {\it Phys.Rev.Lett.} {\bf 59}, (1987) 1873;
M.Henneaux and C.Teitelboim, in {\it Proc. Quantum Mechanics of Fundamental
Systems 2, Santiago (1987) 79}; {\it Phys.Lett.} {\bf 206B} (1988) 650;
A.Tseytlin, {\it Phys.Lett.} {\bf 242B} (1990) 163; {\it Nucl.Phys.} {\bf B350}
(1991) 395; J.H.Schwarz and A.Sen, {\it Nucl.Phys.} {\bf B411} (1994) 35.
\bibitem [3] {kn:xx}B.McClain, Y.S.Wu and F.Yu, {\it Nucl.Phys.} {\bf B343} (1990) 689;
C.Wotzacek, {\it Phys.Rev.Lett.} {\bf 66} (1991) 129 ;
I.Martin and A.Restuccia, {\it Phys.Lett.} {\bf 323B} (1994) 311 ;
F.P.Devecchi and M.Henneaux, {\it Phys.Rev.} {\bf D45} (1996) 1606 ;
I.Bengtsson and A.Kleppe, {\it Int.J.Mod.Phys.} {\bf A12} (1997) 3397 ;
N.Berkovits, {\it Phys.Lett.} {\bf 388B} (1996) 743 ; {\it Phys.Lett.}
{\bf 395B} (1997) 28 ; {\it Phys.Lett.} {\bf 398B} (1997) 79.
\bibitem [4] {kn:xx}W.Seigel, {\it Nucl.Phys.} {\bf B238} (1984) 307 ;
P.Pasti, D.Sorokin and M.Tonin, {\it Phys.Lett.} {\bf 352B} (1995) 59 ;
{\it Phys.Rev.} {\bf D52} (1995) R4277 ; P.Pasti, D.Sorokin and M.Tonin,
in {\it Leuven Notes in Mathematical and Theoretical Physics} (Leuven Univ.Press,
Series B Vol.6, (1996) p.167) ; {\it Phys.Rev} {\bf D55}  (1997) 6292.
\bibitem [5] {kn:xx}A.Maznytsia, C.R.Preitschopf and D.Sorokin, {\it hep-th/9805110, hep-th/9808049}.
\bibitem [6] {kn:xx}S.Deser, A.Gomberoff, M.Henneaux and C.Teitelboim, {\it Phys.Lett.} {\bf 400B} (1997) 80.
\bibitem [7] {kn:xx}R.Medina and N.Berkovits, {\it Phys.Rev.} {\bf D56} (1997) 6388.
\bibitem [8] {kn:xx}P.Pasti, D.Sorokin and M.Tonin, {\it Phys.Lett.} {\bf 398B} (1997) 41 ; I.Bandos,
K.Lechner, A.Nurmagambetov, P.Pasti, D.Sorokin and M.Tonin, {\it Phys.Rev.Lett.}{\bf 78} (1997) 432;
{\it Phys.Lett.} {\bf 408B} (1997) 135 ; I.Bandos, M.Cederwall and D.Sorokin, {\it Nucl.Phys.} {\bf B522} (1998) 214 ;
G.Dall'Agata, K.Lechner and D. Sorokin, {\it Class.Quant.Grav.} {\bf 14} (1997) L195 ;
G.Dall'Agata, K.Lechner and M.Tonin, {\it hep-th/9806140}.
\bibitem [9] {kn:xx}G.Dall'Agata and K.Lechner, {\it Nucl.Phys.} {\bf B511} (1998) 326 ;
G.Dall'Agata, K.Lechner and M.Tonin, {\it Nucl.Phys.} {\bf B512} (1998) 179 ; A.Nurmagambetov, {\it hep-th/9804157}.
\bibitem [10] {kn:xx}M.K.Gaillard and B.Zumino,{\it Nucl. Phys.} {\bf B193},
(1981) 221, {\it hep-th/9705226,hep-th/9705226} ; G.W.Gibbons and D.A.Rasheed,
{\it Nucl.Phys.}{\bf B454} (1995) 185,{\it Phys.Lett.}{\bf B365} (1996) 46 ;
Y.Igarashi,K.Itoh and K.Kamimura, {\it hep-th/9806160, hep-th/9806161}.
\bibitem [11] {kn:xx}A.S.Schwarz, in {\it Topology for Physicists} (Springer Verlag,Berlin Heidelberg,1994) ;
P.A.M. Dirac, {\it Proc.R.Soc.} {\bf A133} (1931) 60 ; G.t'Hooft, {\it Nucl. Phys.}
{\bf B79} (1974) 276 ; A.M.Polyakov {\it JETP Lett.} {\bf 20} (1974) 194.
\bibitem [12] {kn:xx}H.Goldstein, {\it Classical Mechanics, Second Edition} (Addison-Wesley, New York, 1980 p.545-600).
\bibitem [13] {kn:xx} G.t'Hooft, {\it Dimensional Reduction in Quantum Gravity,gr-qc/9310006};
L.Susskind, {\it Phys.Rev.} {\bf D49} (1994) 1912;
J.Maldacena, {\it Adv.Theor.Math.Phys.} {\bf 2} (1998) 231;
E.Witten, {\it Adv.Theor.Math.Phys.} {\bf 2} (1998) 253;
L.Susskind and E.Witten,{\it The Holographic bound in Anti-de Sitter Space, hep-th/9805114};
O.Aharony, S.S.Gubser,J.Maldacena,H.Ooguri and Y.Oz, {\it Large N Field Theories,
String Theory and Gravity, hep-th/9905111}.


\end{thebibliography}
\end{document}